\documentclass{PoS}

\usepackage{lineno}

\usepackage{csquotes}
\title{ Synergy between Art and Science: Collaboration at the South Pole.  }

\ShortTitle{ Synergy between Art and Science: Collaboration at the South Pole}

\author{
The IceCube Collaboration\footnote{For collaboration list, see PoS(ICRC2019) 1177.}\\
{\itshape \href{http://icecube.wisc.edu/collaboration/authors/icrc19_icecube}{http://icecube.wisc.edu/collaboration/authors/icrc19\_icecube}}\\
E-mail: \email{dfortescue@cca.edu, gdewasse@icecube.wisc.edu}
}

\abstract{
We present the result of a cross-disciplinary collaboration between Prof. Donald Fortescue of the California College of the Arts in San Francisco and the Dr. Gwenhael de Wasseige of the IceCube Collaboration. The work presented was initiated during Fortescue's US National Science Foundation funded Antarctic Artists and Writers Fellowship at the South Pole in the austral summer of 2016/17. One outcome of this collaboration is the video work \textit{Axis Mundi} - a timelapse movie captured during 24 hours at the South Pole, combined with a simultaneous sampling of IceCube data transduced into sound. \textit{Axis Mundi} captures the rotation of the Earth in space, the transient motions of the atmosphere, and the passage of subatomic particles through the polar ice, to provide a means for us to physically engage with these phenomena. We detail how both the timelapse and the transduction of atmospheric muon data have been realized and discuss the benefits of such a collaboration.
\\

\vspace{4mm}
{\bfseries Corresponding authors:}
\speaker{Prof. Donald Fortescue}$^{1}$,  Gwenhael de Wasseige$^{2}$\\
{$^{1}$ \itshape California College of the Arts, San Francisco }
{$^{2}$ \itshape Laboratoire APC, Paris-Diderot, France}

}

\FullConference{36th International Cosmic Ray Conference -ICRC2019-\\
		July 24th - August 1st, 2019\\
		Madison, WI, U.S.A.}

\begin{document}

\section{Introduction}\label{intro}
Fortescue and de Wasseige met in Antarctica during the austral summer of 2016/17. Fortescue was conducting research at the South Pole in collaboration with the IceCube Neutrino Observatory with the support of an Artists and Writers Fellowship through the US National Science Foundation. De Wasseige was working directly with the IceCube collaboration deploying new sensors to augment the modelling of ice surface conditions for the Observatory. De Wasseige and Fortescue collaborated on developing the sound work, \textit{86 Strings \#1}, involving the sonification of data from muon interactions in the IceCube array.

\section{Process}\label{process}

\subsection{IceCube}
The IceCube Neutrino Observatory at the South Pole is the product of a multinational collaboration~\cite{icecube}.  It was built primarily to detect and analyze high-energy neutrinos from cosmological sources. However, many other observational capacities have been realized since its completion in 2011 - including fundamental neutrino physics, cosmic ray physics, and solar dynamics~\cite{hese, sterile}. 

The IceCube detector array occupies a cubic kilometer of ice and is buried over 1.5 km below the ice surface. The array consists of 5,160 photo-sensitive Digital Optical Modules (DOMs) arrayed on eighty-six 2.5 km long, vertical `strings' that have been lowered into deep holes drilled into the ice and then frozen in place. The majority of the strings are in a hexagonal grid spaced 125 meters apart and each string holds 60 DOMs, with most spaced 17 meters apart vertically. 

This huge volume of ice (c.1 million cubic meters) captures several hundred neutrino interactions every day (most of them being of atmospheric origin), together with hundreds of millions of atmospheric muons. These muons constitute the dominant background for most of the analyses carried out within the IceCube Collaboration (e.g., \cite{hese,sterile,solar}), while for others they are the studied signal\cite{sun-shadow}.

\begin{figure}[th!]
    \centering
    \includegraphics[width=0.85\textwidth]{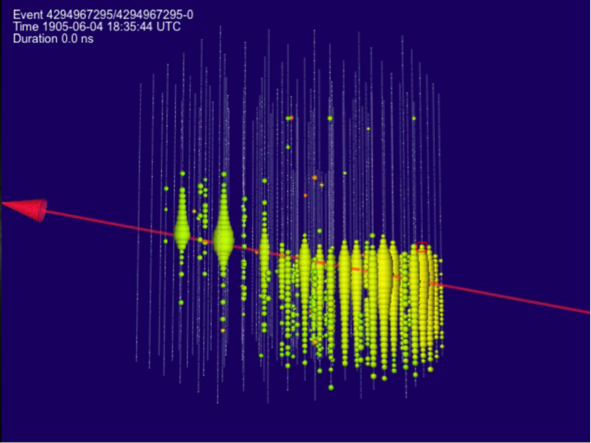}
    \caption{IceCube data visualisation in Steamshovel. The red arrow indicates the estimated pathway of a superluminal muon through the IceCube array. Credits: IceCube Collaboration. }\label{event}
\end{figure}

A neutrino interacting with an ice nucleon may result in a  relativistic muon that generates Cherenkov photons, which are detected by the DOMs. These photons are transduced by the DOMs into electrical impulses, converted to data streams and then sorted by computer algorithms at the IceCube Lab to identify potentially significant signals from the ever-present sea of noise. These particle interactions are visualized by IceCube through their purpose-built Steamshovel software program. Figure~\ref{event} shows a screen capture from a  Steamshovel animation of a single event in the ice. The image shows the 86 strings of the array and the individual DOMs arranged like beads on each string. One of this project's aims was to understand what this data could `sound' like rather than `look' like, if it was transduced to sound instead of pixels on a screen. 

The number of strings in the array and the fact that scientists call the long cables and attached instrumentation `strings' led Fortescue to think of IceCube as an enormous stringed instrument. This conceptualization led directly to the idea of mapping the 86 strings of the array onto the 88 keys of a grand piano, and envisioning the photon `hits' on individual DOMs as strikes on the strings of the piano. 

While working at the South Pole, Fortescue discussed this notion with de Wasseige and Martin Rongen (both of the IceCube Collaboration) who enthusiastically lent their expertise to help make this a reality. The collaboration between Fortescue and de Wasseige, begun when they met in Antarctica, has continued over the following two years as they developed an audio work from this approach. The resulting audio work has been titled \textit{86 Strings \#1}.

\subsection{Data sonification}

Two terms used by both scientists and sound artists working with converting data to sound are `audify' and `sonify'. Audification is the making audible of an inaudible sound through amplification or through transposing frequencies into the range that humans can hear - this is the process that makes whale song audible to us. Audification is essentially sound transcription. Sonification is the process of rendering other forms of data (electromagnetic, particle flux or gravitational wave signals, for example) into sound - a classic example is the Geiger counter. Sonification is the transduction of data into into audible sound. \textit{86 Strings \#1} involves the sonification of digital signals derived from the detection of photons deep in the polar ice into sounds we can hear.

There are innumerable ways that the IceCube data could be transduced into sound. Each requires the allocation of sound frequency, volume, duration and timing values to map the event rate, deposited energy, photon flux and location values in the data. Some mappings are readily suggested by the shared characteristics of sound and light waves (such as frequency, intensity, and duration). Others (such as which events are sampled, the sampling rate and playback speed) need to be selected to both reflect the underlying physics and to satisfy aesthetic considerations (that is, to ensure the resulting work will be engaging). In the case of \textit{86 Strings \#1} we assigned the values of the timing, intensity and duration of the DOM signals to timing, loudness and sustain on each struck note. 

\begin{figure}[th!]
    \centering
    \includegraphics[width=0.85\textwidth]{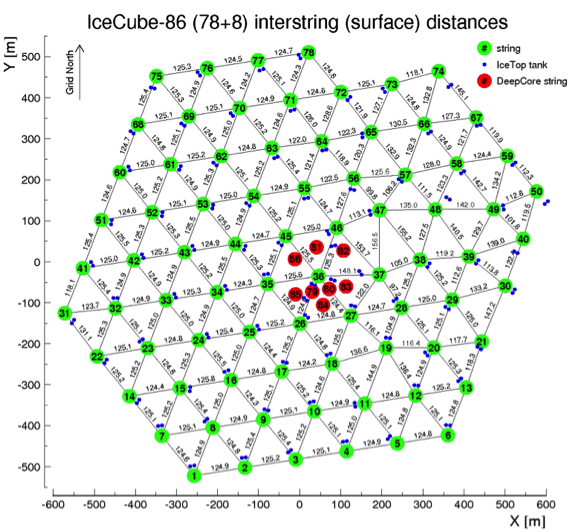}
    \caption{Map of the current arrangement of the 86 strings of DOMs in the IceCube array. Credits: IceCube Collaboration.  }\label{map}
\end{figure}

The strings in the IceCube array have already been allocated numbers (see Figure~\ref{map}). We decided to follow this numbering system by allocating the 1st string to the lowest note on the grand piano (A0) and then allocating each note in sequence after that. The choice to assign a particular note to a particular IceCube string highlights the physical movement of muons through the ice - interactions which result in horizontal muon paths result in distinctive glissandos and vertical paths result in repetitive strikes of the same or closely tuned notes. 

\begin{figure}[th!]
    \centering
    \includegraphics[width=0.95\textwidth]{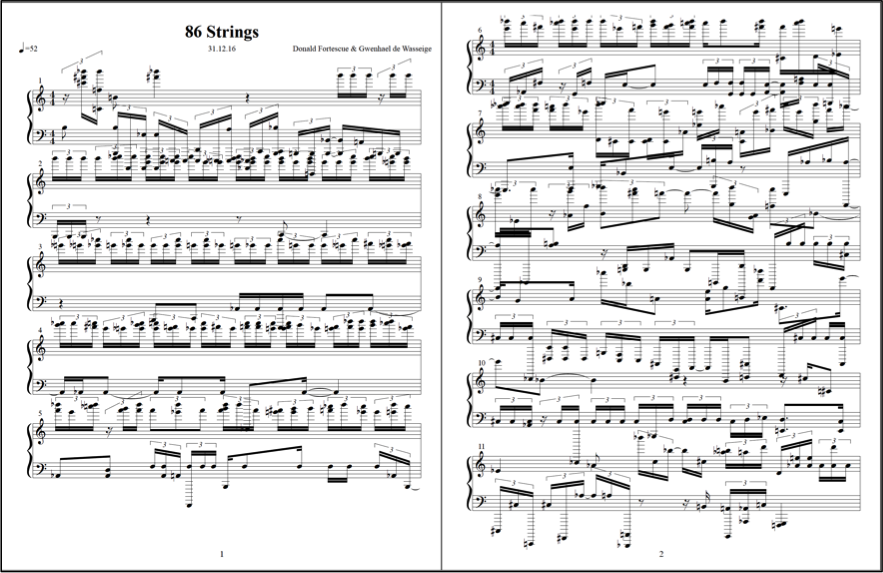}
    \caption{\textit{86 Strings \#1} transcribed to Western musical notation.}\label{partition}
\end{figure}

The conversion of IceCube data to sound is relatively straightforward. IceCube has developed software based on the versatile and widely used, open source software language Python to analyze data sets. This allows ready access to the existing open-source Python libraries, one of which facilitates output to the MIDI format - the international standard software system for digital musical instrumentation. A consequence of this software synergy is that IceCube data can be readily output to MIDI files, which can then be used to activate a wide range of electronic instruments - including grand piano. Using standard MIDI software the sound can also be directly transcribed to Western musical notation (see Figure~\ref{partition}) - a transcription of the transduction of IceCube data. 

\subsection{Data selection}

IceCube detects one neutrino every 6 minutes on average and 3000 muons per second.  The critical decision as to which events are selected and the tempo at which they play can be aesthetically determined or constrained by other conditions.

Fortescue decided to pair \textit{86 Strings \#1} with a 24 video time-lapse video of the ice surface above the array to create the combined audio-visual work \textit{Axis Mundi}. This work captures the rotation of the Earth in space and the transient motions of the atmosphere, together with the synchronous passage of subatomic particles through the polar ice. This pairing provided an appropriate constraint for the selection and timing of events for the sound work. \textit{86 Strings \#1} is a sampling of events occurring on the same day that was sampled by the timelapse video in \textit{Axis Mundi}. As \textit{Axis Mundi} is focused on the apparent movement of the sun it was decided to sample only events coming from the direction of the sun for the accompanying sound work. We sampled a single muon event at the beginning of each hour. Each event was transduced to sound according to our specifically designed algorithm, which resulted in 10-20 seconds of sound for each event. The starting points of these individual events were then evenly spaced across the duration of the video timelapse. Only 24 events were sampled for the entire soundwork - one from the beginning of each hour. Each event only lasts microseconds in real time and so each has been slowed down by a factor of millions as part of the sonification process.

\textit{86 Strings \#1} is combined with \textit{Axis Mundi}, which is the same 24 hour time period's worth of visual data from the ice surface -  marked by the presence of Fortescue's sculptural work \textit{Instrument (90$^\circ$S)}.  Both timelapses last 4 minutes and 48 seconds and are precisely synchronised. \textit{Axis Mundi} can be viewed at \textcolor{blue}{\underline{https://vimeo.com/314347886}}. A single still from the video is shown in Figure 4.

\begin{figure}[th!]
    \centering
    \includegraphics[width=0.85\textwidth]{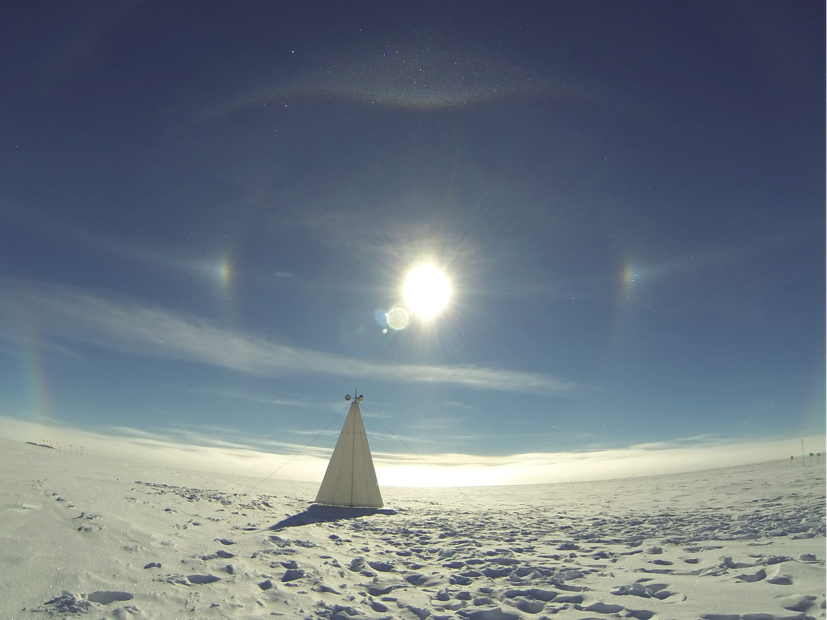}
    \caption{Video still from \textit{Axis Mundi}. 2017. HD (1080p) video with sound.}\label{axismundi}
\end{figure}

\section{Outcomes and Perspectives}

\textit{Axis Mundi} has been presented to both science and art audiences in the Australia and the USA and has bought awareness of the research of the IceCube collaboration to new audiences. It provides a window onto leading astrophysical research, which can be difficult for lay audiences to engage with.

\subsection{Objective approaches to art making}
\textit{Axis Mundi} is a compelling and intriguing work. \textit{86 Strings \#1} provides an appropriately dramatic accompaniment to the surreal phenomenon of the unsetting sun circling at seemingly fixed altitude in the  South Polar sky. 

Many viewers have assumed that \textit{86 Strings \#1} was `composed' specifically for the video. In a sense it was, but the individual notes and their timing were not selected by the work's authors. Instead we established an objective set of constraints on the data set selected and the way this data was then transduced into sound. This is analogous to the way any scientific study uses a blind procedure: developing the analysis on `background-only' data, selecting relevant 'signal' data and then processing that data for analysis. 

Such an objective approach to art making might at first glance seem to deny the creativity of the artist. However, this approach has a strong precedent in contemporary art. An objective methodology where distinct rules are established to constrain artists choices was first established by the Minimalist and Conceptual artists of the 60's and 70's who were striving to reduce artistic agency in reaction to the dominant paradigm of Abstract Expressionism. The renowned conceptual artist Sol LeWitt explained  in his Paragraphs on Conceptual Art~\cite{lewitt} - 
	\begin{displayquote}\textit{To work with a plan that is preset is one way of avoiding subjectivity. It also obviates the necessity of designing each work in turn. The plan would design the work. Some plans would require millions of variations, and some a limited number, but both are finite. ... In each case, ... the artist would select the basic form and rules that would govern the solution of the problem. After that the fewer decisions made in the course of completing the work, the better. This eliminates the arbitrary, the capricious, and the subjective as much as possible.} \end{displayquote}
This objective approach provides a strong link between the methodological practice of science and art and can help delineate the required conditions for effective art/science collaborations.

\subsection{Value of sonification}
The potential for original approaches to sonification of data to be useful in scientific analysis has proven to be a key incentive for continuing this collaboration beyond our initial efforts. 
We readily discern sound as occurring in three dimensions so it is anticipated that listening to neutrino interactions could be more intuitively informative to scientists than looking at animated renderings on a 2D screen. 

In her analysis of the visualization methods employed by the NASA Rover team, the sociologist of science, Janet Vertesi coined the term `drawing as' to describe the practice of `theory-laden representation' - creating imagery that embodies ways of seeing and ways of thinking and foregrounds the process involved in making the images. To draw also connotes \textit{to pull or guide, to reveal and conceal, to work with and around material objects, to produce new configurations of space and movement}~\cite{vertesi}.  It is a physical and embodied process. By analogy with `drawing as', \textit{86 Strings \#1} can be considered as a form of `listening as'. And like `drawing as', `listening as' can be seen as both a `theory-laden representation' and a physical and embodied process~\cite{vertesi-2}.  

It is possible within \textit{86 Strings \#1 }to hear certain distinct characteristics of the muons  passing through the ice. As the Sun travels through the sky, one can hear the change in the incoming direction of the recorded muons, through the changes in the tone.  The path followed by the muons when crossing IceCube can similarly be reconstructed through the variation of tone within the event itself. Furthermore, the stochastic loss of energy characteristic to muons is clearly modeled with the variation of the tempo representing more or less photons recorded as the muons travel through IceCube.

\subsection{Value of Collaboration}
One of the most significant outcomes of this project has been the development of a lasting collaboration between Fortescue and de Wasseige. The success of the collaboration comes from our mutual interest in each other's research, our proven ability to work effectively together (initially in the challenging conditions of Antarctica and subsequently by email and skype from opposite time zones in Europe and the West Coast of the US), and our desire to reach new audiences with our interdisciplinary research and its outcomes. This fruitful collaboration is continuing through a project we are developing with KM3NeT~\cite{km3net} in the Mediterranean Sea. 

\section{Conclusion}
This fruitful collaboration has generated an audio visual work which exemplifies new approaches to both sonification of astrophysical data and to art making. We are continuing to explore these and other new approaches in a project in development with KM3NeT.

\section*{Acknowledgements}
Fortescue's research in Antarctica, in the austral summer of 2016/2017, was supported by the United States National Science Foundation (NSF) through an Antarctic Artists and Writers Fellowship, by the School of Art and Design at the Australian National University and by an Australian Government Research Training Program Scholarship. Particular thanks go to the scientists and staff of the IceCube Neutrino Observatory for their advice and assistance and for hosting Fortescue and de Wasseige at the South Pole. The data used in this research was generously provided by and used with the permission of the IceCube Collaboration. This project would never have gotten off the ground without the direct and generous support of Dr. Jim Madsen (Associate Director for Education \& Outreach at the IceCube Neutrino Observatory).


%

\end{document}